\documentclass[4apaper,11pt]{article}
\usepackage{amssymb,amsfonts,amsmath, psfrag,eepic}
\makeatletter \@addtoreset{equation}{section}
\@addtoreset{theo}{section} \makeatother

\def\qed{\hfill \rule{4pt}{7pt}}

\title{
\bf The neighbor-scattering number can be computed in polynomial
time for interval graphs\footnote{ Supported by NSFC. } }
\author{
\small  Fengwei Li and Xueliang Li\\
[2mm]
\small Center for Combinatorics and LPMC \\
\small Nankai University\\
\small Tianjin 300071, P.R. China \\
\small fengwei.li@eyou.com, x.li@eyou.com\\
}
\date{ }
\begin{document}
\maketitle \vskip5mm
\begin{abstract} {Neighbor-scattering number is a useful measure for graph
vulnerability. For some special kinds of graphs, explicit formulas
are given for this number. However, for general graphs it is shown
that to compute this number is NP-complete. In this paper, we prove
that for interval graphs this number can be computed in polynomial
time.}
\vspace{0.3cm}\\
{\bf Keyworks:} neighbor-scattering number, interval graph,
consecutive clique arrangement.
\end{abstract}

\section{Introduction}
Throughout this paper, we use Bondy and Murty $[1]$ for terminology
and notations not defined here and consider finite simple undirected
graphs only. The vertex set of a graph $G$ is denoted by $V$ and the
edge set of $G$ is denoted by $E$. We always denote the number of
vertices of $G$ by $n$ and the number of edges of $G$ by $m$. By
$\omega(G)$ we denote the number of components of $G$. $deg(v)$
denotes the degree of a vertex $v$ in $G$. If $S$ is a vertex subset
of $V$, we use $G[S]$ to denote the subgraph of $G$ induced by $S$.

The scattering number of a graph was introduced by Jung [9] as an
alternative measure of the vulnerability of graphs to disruption
caused by the removal of vertices.

In $[6,7,8]$ Gunther and Hartnell introduced the idea of modeling a
spy network by a graph whose vertices represent the agents and whose
edges represent lines of communication. Clearly, if a spy is
discovered or arrested, the espionage agency can no longer trust any
of the spies with whom he or she was in direct communication, and so
the betrayed agents become effectively useless to the network as a
whole. Such betrayals are clearly equivalent to the removal of the
closed neighborhood of $v$ in the modeling graph, where $v$ is the
vertex representing the particular agent who has been subverted.

Therefore, instead of considering the scattering number of a
communication network, we discuss the (vertex) neighbor-scattering
number of graphs - disruption caused by the removal of vertices and
their adjacent vertices.

Let $G=(V,E)$ be a graph and $u$ a vertex in $G$. The {\it open
neighborhood} of $u$ is $N(u)=\{v\in V(G)|(u,v)\in E(G)\}$, and the
{\it closed neighborhood} of u is $N[u]=\{u\}\cup N(u)$. We define
analogously for any $S\subseteq V(G)$ the open neighborhood $N(S)=
\cup_{u\in S}N(u)$ and the closed neighborhood $N[S] = \cup_{u\in
S}N[u]$. A vertex $u\in V(G)$ is said to be {\it subverted} when the
closed neighborhood $N[u]$ is deleted from $G$. A {\it vertex
subversion strategy} of $G$, $X$, is a set of vertices whose closed
neighborhood is deleted from $G$. The survival-subgraph, $G/X$, is
defined to be the subgraph left after the subversion strategy $X$ is
applied to $G$, i.e., $G/X = G-N[X]$. $X$ is called a {\it
cut-strategy} of $G$ if the survival subgraph $G/X$ is disconnected,
or a clique, or $\emptyset$.\\
[2mm]{\bf Definition 1.1} ([12]) The {\it(vertex)
neighbor-scattering number} of a graph $G$ is defined as
$$S(G)= max\{\omega(G/X)-|X|: X\ is\ cut-strategy\ of\ G,\ \omega(G/X)\geq 1\},$$
where the maximum is taken over all the cut-strategies of $G$,
$\omega(G/X)$ is the number of connected components in
the graph $G/X$. Especially, we define $S(K_{n})=1$. \\
[2mm]{\bf Definition 1.2}  A cut-strategy $X$ of $G$ is called an
$S$-set of $G$ if $S(G)=\omega(G/X)-|X|$.

In $[11]$, F. Li and X. Li proved that, in general, the problem of
computing the neighbor-scattering number of a graph is NP-complete.
So, it is interesting to compute the neighbor-scattering number of
special graphs, and some results of this kind were obtained in
$[12]$. In Section $3$, we prove that for interval graphs the
neighbor-scattering number can be computed in polynomial time.
Before proving this, in Section $2$, we need to set up relationship
between neighbor-scattering number and minimal cut-strategy of a
graph and give a formula for calculating the neighbor-scattering
number.

\section{ Minimal cut-strategy and neighbor\\-scattering number}

In this section, we characterize the property of minimal
cut-strategy $X$ of a graph $G$ with $\omega(G/X)\geq 1$, and give a
formula to calculate neighbor-scattering number via minimal
cut-strategy $X$ of a graph $G$ with $\omega(G/X)\geq 1$. First, we
give the definition of the minimal cut-strategy of a graph $G$ as follows.\\
[2mm]{\bf Definition 2.1} A subset $X\subset V$ is a {\it
cut-strategy} of a graph $G=(V,E)$ if $G/X$ is disconnected, a
clique, or $\emptyset$. If no proper subset of $X$ is a {\it
cut-strategy} of graph $G$, then $X$ is called a {\it minimal
cut-strategy}
of $G$.\\
[2mm]{\bf Remark.} From the above definition we know that if $X$
is a minimal cut-strategy of graph $G$, then the removal of closed
neighborhood of any vertex set $X'\subset X$ neither disconnects
$G$ nor results in the remaining subgraph being a clique.\\
[2mm]{\bf Lemma 2.1} {\it Let $X=\{v_{1},v_{2},\cdots,v_{t}\}$,
$t\geq 1$, be a {\it cut-strategy} of graph $G$ with
$\omega(G/X)\geq 1$, then $X$ is a minimal {\it cut-strategy} of $G$
if and only if one of the following conditions is satisfied:\\
[2mm]$(a)$ There are at least two different connected components,
say $C_{1}, C_{2}, \cdots,C_{k}$ $(k\geq 2)$, in $G/X$. For every
vertex $v_{i}\in X$ and every connected component $C_{j}$
$(j=1,2,\cdots,k)$ of $G/X$, $v_{i}$ has a neighbor set $B_{ij}$ in
$N(C_{j})$. And if $|X|\geq 2$, for distinct vertices $v_{s}$ and
$v_{t}$ in $X$, neither $B_{si}\subseteq B_{ti}$ nor
$B_{ti}\subseteq B_{si}$ for $C_{i}$ $(i=1,2,\cdots,k)$. For every
vertex $v\in X$, if $v_{j}\in N[v]$ and there doesn't exist any edge
joining $v_{j}$ with any component $C_{i}$ $(i=1,2,\cdots,k)$, then
there exists no edge joining $v$ with other vertex in $X$ if
$|X|\geq 2$. Furthermore, $X$ must be an independent set of $G$ in this case.\\
[2mm]$(b)$ $G/X$ is a maximal clique $C$ and every vertex $v_{i}\in
X$ has a neighbor $B_{i}$ in $N(C)$, and if $|X|\geq 2$, for
distinct vertices $v_{i}$ and $v_{j}$ in $X$, neither
$B_{i}\subseteq B_{j}$ nor $B_{j}\subseteq B_{i}$ for $C$.
Furthermore, for $v\in X$, if $v_{j}\in N[v]$ and there doesn't
exist any edge join $v_{j}$ with this clique, then there exists no
edge joining $v$ with other vertex in $X$ if $|X|\geq 2$. }\\
[2mm]{\bf Proof.} We prove the necessity first. If $X$ is a
minimal cut-strategy of $G$, then $(a)$ or $(b)$ must hold.
We distinguish two cases:\\
[2mm]{\bf Case 1.} If $(a)$ does not hold, we assume there exists a
vertex $v$ in $X$ which does not have any neighbor in the open
neighborhood of one of these components. Without loss of generality,
we assume that for component $C_{i}$ $(i=1,2,\cdots,k-1)$, $v$ has a
neighbor in the open neighborhood of these components but $v$ does
not have any neighbor in the open neighborhood of component $C_{k}$.
It is easy to see that under this condition $X'=X-{v}$ is also a
cut-strategy of $G$ with $\omega(G/X')\geq 2$, a contradiction to
the minimality of $X$. Thus for every vertex $v\in X$ and every
connected component $C_{i}$ $(i=1,2,\cdots,k)$ of $G/X$, $v$ has at
least one neighbor in $N(C_{i})$.

When $|X|\geq 2$, for distinct vertices $v_{s}$ and $v_{t}$ in $X$,
if either $B_{si}\subseteq B_{ti}$ or $B_{ti}\subseteq B_{si}$ for a
same component $C_{i}$ $(1\leq i\leq k)$. Without loss of
generality, we suppose that $B_{s1}\subseteq B_{t1}$ or
$B_{t1}\subseteq B_{s1}$ for $C_{1}$, then it is easily seen that
$X'=X-{v_{s}}$ or $X'=X-{v_{t}}$ is also a cut-strategy of $G$ with
$\omega(G/X')\geq 2$, a contradiction to the minimality of $X$.

When $|X|\geq 2$, if $v_{j}\in N[v]$ and there doesn't exist any
edge joining $v_{j}$ with any component $C_{i}$ $(i=1,2,\cdots,k)$,
then there exists no edge joining $v$ with other vertex in $X$.
Otherwise, if there exists an edge joining $v$ with a vertex $v'\in
X$, then it is easily checked that $X'=X-v$ is a cut-strategy of $G$
with $\omega(G/X')\geq 2$, a contradiction to the minimality of $X$.
So, there exists no edge joining $v$ with other vertex in $X$.

Under this condition, $X$ must be an independent set of $G$,
otherwise, if we have vertices $u, v\in X$ and $(u,v)\in E(G)$, then
$X'=X-v$ is obvious a cut-strategy of $G$. A contradiction to the
minimality of $X$.\\
[2mm]{\bf Case 2.} If $(b)$ does not hold, there must exist a vertex
$v$ in $X$ which does not have any neighbor in the open neighborhood
of the only clique of $C=G/X$. It is obvious that under this
condition there must exist an edge $(v,u)$ joining $v$ with a vertex
$u\in X$, otherwise contradicts the fact that $G$ is connected. It
is easily checked that $X'=X-{v}$ is also a cut-strategy of $G$ with
$\omega(G/X')\geq 1$, a contradiction to the minimality of $X$.

If $|X|\geq 2$, then for distinct vertices $v_{s}$ and $v_{t}$ in
$X$, if either $B_{s}\subseteq B_{t}$ or $B_{t}\subseteq B_{s}$ for
$C$, it is easily seen that $X'=X-{v_{s}}$ or $X'=X-{v_{t}}$ is also
a cut-strategy of $G$ with $\omega(G/X')\geq1$, a contradiction to
the minimality of $X$.

When $|X|\geq 2$, if $v_{j}\in N[v]$ and there doesn't exist any
edge joining $v_{j}$ with clique $C$, then there exists no edge
joining $v$ with other vertex in $X$. Otherwise, if there exists a
vertex $v'\in X$ and $(v,v')\in E(G[X])$, it is easy to see that
$X'=X-v$ is a cut-strategy of $G$ with $\omega(G/X')\geq2$, a
contradiction to the minimality of $X$. So, there exists no edge
joining $v$ with other vertex in $X$.

The proof of the sufficiency proceeds in the following two cases:\\
[2mm]{\bf Case 1.} If $(a)$ holds, then $X$ must be a minimal
cut-strategy of graph $G$. Otherwise, there exists a subset
$X'\subset X$ which is a cut-strategy of graph $G$. Then, for every
vertex $v\in X-X'$, $v$ has a neighbor in each neighborhoods of
these components, and we know that $X$ is an independent set, i.e.,
there exists no edge between $X-X'$ and $X'$, so, the graph $G/X'$
is connected. And under this condition $G/X'$ is not a clique, for
there exists no edge joining $v$ with any components of $G/X'$. This
leads to a contradiction to the hypothesis that $X'$ is a
cut-strategy of graph $G$.\\
[2mm]{\bf Case 2.} If $(b)$ holds, then $X$ must be a minimal
cut-strategy of graph $G$. Otherwise, there exists a subset
$X'\subset X$ which is a cut-strategy of graph $G$. Then, for every
vertex $v\in X-X'$, $v$ has at least one neighbor in the
neighborhood of this clique, so, the graph $G/X'$ is connected, and
under this condition $G/X'$ is not a clique, otherwise contradicts
the fact that for distinct vertices $v_{i}$ and $v_{j}$ in $X$,
neither $B_{i}\subseteq B_{j}$ nor $B_{j}\subseteq B_{i}$ for $C$.
This leads to a contradiction to the hypothesis that $X'$ is a cut-strategy
of graph $G$. Thus the proof is completed. \qed \\
[2mm]{\bf Theorem 2.2} {\it Let $G$ be a noncomplete graph. Then
$$S(G)=max_{X^{\ast}}\{\sum_{i=1}^{k}max\{S(G[C_{i}]),1\}-|X^{\ast}|\} \ \ \ \ \ \ \  (1)$$
where the maximum is taken over all minimal cut-strategies
$X^{\ast}$ of the graph $G$ with $\omega(G/X^{\ast})\geq 1$ and the
$C_{1}, C_{2}, \cdots,C_{k}$ are the connected components of
$G/X^{\ast}$.}\\
{\bf Proof.} First let $X$ be an $S$-set of $G$, i.e.,
$S(G)=\omega(G/X)-|X|$ and $\omega(G/X)\geq 1$. Let $X^{\ast}$ be a
minimal cut-strategy of $G$ with $\omega(G/X^{\ast})\geq 1$ that is
a subset of $X$ and let $C_{1}, C_{2}, \cdots,C_{k}$ be the
connected components of $G/X^{\ast}$. We consider the sets
$X_{i}=X\cap C_{i}$, $i\in \{1,2,\cdots,k\}$. The proof proceeds
in the following two cases:\\
[2mm]{\bf Case 1.} If we assume $X_{i}=\emptyset$, i.e., $X\subset
N[X^{\ast}],$ then we know that $N(X_{i})=\emptyset$ is not a
cut-set of $C_{i}$, i.e., $X_{i}$ is not a cut-strategy of
$G[C_{i}]$. Then, $\omega(C_{i}/X_{i})=1$, hence,
$\omega(C_{i}/X_{i})-|X_{i}|=1$.\\
[2mm]{\bf Case 2.} Now assume $X_{i}\neq\emptyset$. Suppose that
$X_{i}$ is not a cut-strategy of $G[C_{i}]$. Then we have
$\omega(G/(X-X_{i}))=\omega(G/X)$. Furthermore, it is obvious that
$\omega(G/(X-X_{i}))-|X-X_{i}|=\omega(G/X)-|X|+|X_{i}|>\omega(G/X)-|X|=S(G),$
a contradiction to the definition of neighbor-scattering number of graphs.\\
[2mm]\hspace*{6mm}Hence $X\neq\emptyset$ implies that $X_{i}$ is a
cut-strategy of $C_{i}$. Thus,
$S(G[C_{i}])\geq \omega(C_{i}/X_{i})-|X_{i}|$.\\
[2mm]\hspace*{6mm}Summing up the values of
$\omega(C_{i}/X_{i})-|X_{i}|$ over all components $C_{i}$ of
$G/X^{\ast}$ will achieve the value of $\omega(G/X)-|X|=S(G)$. Thus
we have
$S(G)=\omega(G/X)-|X|=\sum_{i=1}^{k}\{\omega(C_{i}/X_{i})-|X_{i}|\}-|X^{\ast}|\leq
\sum_{i=1}^{k}max\{S(G[C_{i}]),1\}-|X^{\ast}|$.

On the other hand, let $X^{\ast}$ be a minimal cut-strategy of $G$
with $\omega(G/X^{\ast})$ $\geq 1$. Furthermore let $C_{1}, C_{2},
\cdots,C_{k}$ be the connected components of $G/X^{\ast}$. Then we
construct a cut-strategy of $G$ such that
$X=X^{\ast}\cup\cup_{i=1}^{k}X_{i}$ with $X_{i}\subset C_{i}$ for
every $i\in \{1,2,\cdots,k\}$. For $i\in \{1,2,\cdots,k\}$ we set
$X_{i}=\emptyset$ if $S(G[C_{i}])\leq 1$. Otherwise if
$S(G[C_{i}])>1$, we choose a cut-strategy $X_{i}$ of $G[C_{i}]$ with
$\omega(C_{i}/X_{i})\geq1$ such that
$S(G[C_{i}])=\omega(C_{i}/X_{i})-|X_{i}|$. Then $X\supset X^{\ast}$
is a cut-strategy of $G$ and we have
$S(G)\geq\omega(G/X)-|X|=\sum_{i=1}^{k}\{\omega(C_{i}/X_{i})-|X_{i}|\}-|X^{\ast}|$.\\
[2mm]\hspace*{6mm}Without loss of generality, let $C_{1}, C_{2},
\cdots,C_{r}$, $0\leq r\leq k$, be the connected components of $G$
with $S(G[C_{i}])\leq 1$. Consequently, $S(G)=\sum_{i=1}^{k}$
$\{\omega(C_{i}/X_{i})-|X_{i}|\}-|X^{\ast}|
=\sum_{i=1}^{r}1+\sum_{i=r+1}^{k}S(G[C_{i}])-|X^{\ast}|=\sum_{i=1}^{k}max$
$\{S(G[C_{i}]),1\}-|X^{\ast}|$.
This completes the proof.\qed \\
[2mm]{\bf Example 1.} Compute the neighbor-scattering number,
$S(G)$, of the graph $G$ given in Figure $1$.\\
[2mm]{\bf Solution.} Using Lemma $2.1$, it is easy to see that in
the graph $G$, vertices $3,4$ form a minimal cut-strategies with two
components in the survival subgraph, vertices $5,6$ form another
minimal cut-strategies with three components in the survival
subgraph, and vertex $2$ forms another minimal cut-strategy with
only one component in the survival subgraph. Then, by the
definition, we know that $S(G)=2$. On the other hand,
$\sum_{i=1}^{k}max\{S(G[C_{i}]),$ $1\}-|X^{\ast}|=0,\ 1$ or $2$, so
$max_{X^{\ast}}\{\sum_{i=1}^{k}max\{S(G[C_{i}]),1\}-|X^{\ast}|\}=2=S(G)$.

\section{Neighbor-scattering number for interval\\ graphs}

Interval graphs are a large class of graphs and important modeling
for useful networks. In this section we try to compute the
neighbor-scattering number for interval graphs, and prove that
neighbor-scattering number can be computed in polynomial time
for interval graphs. First, we give the definition of an interval graph.\\
[2mm]{\bf Definition $3.1$}([5]) An undirected graph $G$ is called
an {\it interval graph} if its vertices can be put into one to one
correspondence with a set of intervals $\ell$ of a linearly ordered
set (like the real line) such that two vertices are connected by an
edge if and only if their corresponding intervals have nonempty
intersection. We call $\ell$ an interval representation
for $G$.\\
[2mm]{\bf Example 2.} In Figure $1$, we give an interval graph $G$
and its interval representation:

\begin{figure}[h,t]
\begin{center}
\setlength{\unitlength}{2mm}
\begin{picture}(60,20)

\put(4,10){\circle*{0.7}} \put(4,11){$1$}\put(4,10){\line(2,1){4}}
\put(8,12){\circle*{0.7}}
\put(8,12.1){\line(0,1){7}}\put(8,19){\circle*{0.7}}\put(8,10){$2$}\put(8,20){$3$}
\put(8,12.1){\line(2,-1){5}} \put(13,9.7){\circle*{0.7}}
\put(13,8){$4$} \put(13,9.7){\line(-1,2){5}}
\put(8,12){\line(1,0){8}}\put(16,12){\circle*{0.7}}\put(16,10){$5$}
\put(8,12){\line(1,1){8}}\put(16,20){\circle*{0.7}}\put(16,21){$6$}
\put(16,20){\line(0,-1){8}} \put(13,9.7){\line(2,7){3}}
\put(13,9.7){\line(4,3){3}} \put(16,12){\line(2,1){4}}
\put(20,14){\circle*{0.7}}\put(19.4,12){$7$}\put(20,14){\line(2,-1){4}}
\put(24,12){\circle*{0.7}}\put(24,12.5){$8$}

\put(31,16){\line(1,0){4}}\put(36,16){\line(1,0){9}}
\put(31,13){\line(1,0){14}}\put(42,10){\line(1,0){7}}
\put(47,13){\line(1,0){8}}\put(51,10){\line(1,0){4}}
\put(36,19){\line(1,0){4}}\put(41,19){\line(1,0){4}}
\put(33,16.5){$I_{1}$}\put(39.3,16.5){$I_{4}$}\put(37,13.5){$I_{2}$}
\put(49.4,13.5){$I_{7}$}\put(45,10.5){$I_{5}$}\put(52.3,10.5){$I_{8}$}
\put(37.3,19.5){$I_{3}$}\put(42.3,19.5){$I_{6}$}

\end{picture}
\end{center}
\caption{An interval graph $G$ and an interval representation for
it}\label{1em}
\end{figure}
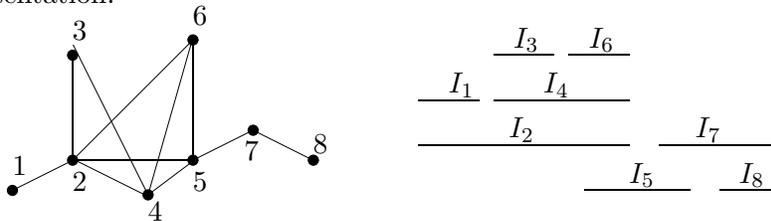

Interval graphs are a well-known family of perfect graphs $[5]$ with
plenty of nice structural properties. The following
characterizations were given by Gilmore and Hoffman $[4]$. \\
[2mm]{\bf Lemma 3.1}([4]) {\it Any induced subgraph of an interval
graph is an interval graph.}\\
[2mm]{\bf Lemma 3.2}(Booth and Leuker [1976])([2]) {\it Interval
graphs can be recognized in $O(m+n)$ time.}\\
[2mm]{\bf Lemma 3.3}(Fulkerson and Gross [1965])([3]) {\it A
triangulated graph on $n$ vertices has at most $n$ maximal
cliques, with equality if and only if the graph has no edges.}\\
[2mm]{\bf Lemma 3.4}([4]) {\it A graph $G$ is an interval graph if
and only if the maximal cliques of $G$ can be linearly ordered, such
that, for every vertex $v$ of $G$, the maximal cliques containing $v$
occur consecutively.}\\
[2mm]Such a linear ordering of the maximal cliques of an interval
graph is said to be a {\it consecutive clique arrangement}. Notice
that interval graphs are triangulated graphs, and by Lemma $3.3$ we
know that an interval graph with $n$ vertices has at most $n$
maximal cliques $[3]$. Booth and Lueker $[2]$ give a linear time
recognition algorithm for interval graphs and the algorithm also
computes a consecutive clique arrangement of the input graph if it
is an interval graph.

Using Lemma $3.1$, we can easily identify the minimal cut-strategy
of an interval graph $G$. And it is easy to see that any minimal
cut-strategy of an interval graph $G$ consists of only one vertex.
When there exist at least three maximal cliques in $G$, if we assume
that vertex $v$ is a minimal cut-strategy of $G$ with
$\omega(G/v)=1$, i.e., $G/v$ is a clique, then by Theorem $2.2$, we
know that $v$ contributes zero to $(1)$. And under this condition,
we can easily find a minimal cut-strategy $u$ with $\omega(G/u)\geq
2$ and it is easily checked that
$\{\sum_{i=1}^{k}max\{S(G[C_{i}]),1\}-|u|\}>\{max\{S(G/v),1\}-|v|\}=0$.
So, when there are at least three maximal cliques in $G$, we only
consider the minimal cut-strategy $v$ with $\omega(G/v)\geq 2$.\\
[2mm]{\bf Theorem 3.5} {\it Let $G$ be an interval graph and let
$A_{i}$, $1\leq i\leq 2$, be a consecutive clique arrangement
of $G$. Then, $S(G)=0$ }\\
[2mm]{\bf Proof.} Under this condition the minimal cut-strategy, say
$v$, of $G$ with $\omega(G/v)=1$ consists of vertex $v\in X=\{v:
v\in A_{1}-S_{1}\ and \ N(v)\cap (A_{2}-S_{1})=\emptyset, or\ v\in
A_{2}-S_{1}\ and\ N(v)\cap(A_{1}-S_{1})=\emptyset \}$. Therefore, by
Theorem $2.2$ we know that
$S(G)=max_{X^{\ast}}\{\sum_{i=1}^{k}max\{S(G[C_{i}]),1\}-|X^{\ast}|\}
=max_{v}\{max\{S(G/v),1\}-|v|\}=0$. \qed \\
[2mm]{\bf Theorem 3.6} {\it Let $G$ be an interval graph and let
$A_{1},A_{2},A_{3}$ be a consecutive clique arrangement of $G$. Then
\[S(G)=\left\{\begin{array}{ll}
1, \ \ if\ there\ exists\ vertex\ v\in A_{i}-(S_{1}\cup
S_{2})-(A_{j}\cup A_{k}),\
 and\ \\ \ \ \ \ \ v\ is\ adjacent\ to\ all\ vertices\ in\ S_{1}\cup S_{2}, i\neq j\neq k\in \{1,2,3\}\\
0, \ \ otherwise \\
\end{array}
\right. \] }\\
[2mm]{\bf Proof.} If there exists a vertex $v\in A_{i}-(S_{1}\cup
S_{2})-(A_{j}\cup A_{k})$, $i\neq j\neq k\in \{1,2,3\}$, such that
it is adjacent to all vertices in $S_{1}\cup S_{2}$, then it is
obvious that $v$ is a minimal cut-strategy of $G$ with
$\omega(G/v)=2$ and the components of $G/v$ are all cliques, thus
$\{max\{S(G/v),1\}-|v|\}=1$. Otherwise, if there is no vertex $v\in
A_{i}-(S_{1}\cup S_{2})-(A_{j}\cup A_{k})$ such that it is adjacent
to all vertices in $S_{1}\cup S_{2}$, then any vertex $v\in
(A_{1}\cup A_{2}\cup A_{3})- (A_{1}\cap A_{2}\cap A_{3})$ is a
minimal cut-strategy of $G$ with $\omega(G/v)=1$, i.e., $G/v$ is a
clique, and then $\{max\{S(G/v),1\}-|v|\}=0$. Hence the proof is
completed. \qed\\
[2mm]{\bf Lemma 3.7} {\it Let $G$ be an interval graph and let
$A_{1}, A_{2}, \cdots, A_{t}$, $t\leq n$, be a consecutive clique
arrangement of $G$. Let $S_{p}=A_{p}\cap A_{p+1}$ for $p\in \{1, 2,
\cdots, t-1\}$. If $t\geq 4$, then the minimal cut-strategy, say
$X$, of $G$ with $\omega(G/X)\geq 2$ consists of vertex $v\in \{v:
2\leq p\leq t-1, v\in A_{p}-(S_{p-1}\cup S_{p}\cup(\cup_{i\neq
p}A_{i})),\ or\ v\in S_{p}-S,\ where\ 2\leq p\leq t-2, S=S_{1}\cup
S_{t-1}\cup (A_{1}\cap A_{2}\cap A_{3})\cup(A_{t-2}\cap A_{t-1}\cap
A_{t}),\ or\ v\in A_{1}-(S_{1}\cup S_{2}\cup(\cup_{i=2}^{t}A_{i}))\
and\ it\ is\ adjacent\ to\ all\ vertices\ in\ S_{1}\cup S_{2}, or\
v\in A_{t}-(S_{t-2}\cup S_{t-1}\cup(\cup_{i=1}^{t-1}A_{i}))\ and\
it\ is\ adjacent\ to\ all\ vertices\ in\ S_{t-2}\cup S_{t-1}\}$} if
there exists no $S_{i}$ and $S_{j}$, $i\neq j$, such that
$S_{i}\subseteq S_{j}$. Otherwise, if there exist $S_{i}$ and
$S_{j}$, $i\neq j$, such that $S_{i}\subseteq S_{j}$, then $v\in\{v:
1\leq p\leq t, v\in A_{p}-S_{j},\ and\ it\ is\
adjacent\ to\ all\ vertices\ in\ S_{j}$\}.\\
[2mm]{\bf Proof.} By Lemmas $2.1$ and $3.4$, it is easily checked
that this Lemma holds.\qed

From above we know that an interval graph $G=(V,E)$ on $n$
vertices has at most $n$ minimal cut-strategies.\\
[2mm]{\bf Definition 3.2}([10]) {\it Let $G$ be an interval graph
with consecutive clique arrangement $A_{1}, A_{2}, \cdots, A_{t}$.
We define $A_{0}=A_{t+1}=\emptyset$. For all $l, r$ with $1\leq
l\leq r\leq t$ we define
$\mathcal{P}(l,r)=(\cup_{i=l}^{r}A_{i})-(A_{l-1}\cup A_{r+1})$. A
set $\mathcal{P}(l,r)$, $1\leq l\leq r\leq t$, is said to be a piece
of $G$ if $\mathcal{P}(l,r)\neq\emptyset$ and $G[\mathcal{P}(l,r)]$
is connected. Furthermore, $V=\mathcal{P}(1,t)$ is a piece of $G$
(even if $G$ is disconnected).}\\
[2mm]{\bf Remark.} It is obvious that cliques in
$\mathcal{P}(l,r)$ are listed in the same order as that they are
listed in graph $G$.\\
[2mm]{\bf Lemma 3.8} {\it Let $X$ be a minimal cut-strategy of
connected subgraph $G[\mathcal{P}(l,r)]$, $1\leq l\leq r\leq t$ with
$\omega(G[\mathcal{P}(l,r)]/X)\geq1$, especially, when
$G[\mathcal{P}(l,r)]$, $1\leq l\leq r\leq t$, has at least four
cliques, $\omega(G[\mathcal{P}(l,r)]/X)\geq 2$. Then there exists a
minimal cut-strategy $X'$ of $G$, such that $X=X'\cap
\mathcal{P}(l,r)=X'-(A_{l-1}\cup A_{r+1})$. Moreover, every
connected component of $G[\mathcal{P}(l,r)/X']$ is
a piece of $G$. }\\
[2mm]{\bf Proof.} By lemma $3.1$, we know that piece
$\mathcal{P}(l,r)$ is an interval graph. And it is obvious that
the linear arrangement $A_{l}-(A_{l-1}\cup A_{r+1})$,
$A_{l+1}-(A_{l-1}\cup A_{r+1})$, $\cdots$, $A_{r}-(A_{l-1}\cup
A_{r+1})$ has all properties of a consecutive clique arrangement
for $\mathcal{P}(l,r)$, except that cliques may occur more than
once. We distinguish three cases:\\
[2mm]{\bf Case 1.} If $\mathcal{P}(l,r)$ has two maximal cliques,
we let $A_{1},A_{2}$ denote these two cliques.\\
Then applying Lemma $3.4$ to $\mathcal{P}(l,r)$ implies that all
minimal cut-strategies of $\mathcal{P}(l,r)$ with
$\omega(G[\mathcal{P}(l,r)]/X)=1$ are sets of the form:\\
When $l\neq 1$ and $r\neq t$, $X'-(A_{l-1}\cup A_{r+1})=\{v: v\in
A_{1}-S_{1}-(A_{l-1}\cup A_{r+1})\ and \ N(v)\cap
(A_{2}-S_{1}-(A_{l-1}\cup A_{r+1}))=\emptyset, or\ v\in
A_{2}-S_{1}-(A_{l-1}\cup A_{r+1})\ and\
N(v)\cap(A_{1}-S_{1}-(A_{l-1}\cup A_{r+1}))=\emptyset \}$.
Especially, when $l=1$, then $X'-(A_{l-1}\cup A_{r+1})=\{v: v\in
A_{2}-S_{1}-(A_{l-1}\cup A_{r+1})\ and \ N(v)\cap
(A_{1}-S_{1}-(A_{l-1}\cup A_{r+1}))=\emptyset\}$. When $r=t$, then
$X'-(A_{l-1}\cup A_{r+1})=\{v: v\in A_{1}-S_{1}-(A_{l-1}\cup
A_{r+1})\ and \ N(v)\cap
(A_{2}-S_{1}-(A_{l-1}\cup A_{r+1}))=\emptyset\}$.\\
[2mm]{\bf Case 2.} If $\mathcal{P}(l,r)$ has three maximal cliques, say $A_{1}, A_{2}, A_{3}$.\\
By applying Lemma $3.6$ to $\mathcal{P}(l,r)$ we get all minimal
cut-strategies of $\mathcal{P}(l,r)$ with
$\omega(G[\mathcal{P}(l,r)]/X)\geq1$.\\
[2mm]{\bf Subcase 2.1} If there exists a vertex $v\in
A_{i}-(S_{l+1}\cup S_{l+2})-(A_{j}\cup A_{k})-(A_{l-1}\cup
A_{r+1})$, and $v$ is adjacent to all vertices in $S_{l+1}\cup
S_{l+2}, i\neq j\neq k\in \{1,2,3\}$, then the minimal
cut-strategies of $\mathcal{P}(l,r)$ with
$\omega(G[\mathcal{P}(l,r)]/X)=2$ are sets of the form
$X'-(A_{l-1}\cup A_{r+1})=\{v: v\in A_{i}-(S_{1}\cup
S_{2})-(A_{j}\cup A_{k})-(A_{l-1}\cup A_{r+1})$, and $v$ is adjacent
to all vertices in $S_{1}\cup S_{2}-(A_{l-1}\cup
A_{r+1}), i\neq j\neq k\in \{1,2,3\}\}$.\\
[2mm]{\bf Subcase 2.2} If there exists no vertex $v\in
A_{i}-(S_{l+1}\cup S_{l+2})-(A_{j}\cup A_{k})-(A_{l-1}\cup
A_{r+1})$, and $v$ is adjacent to all vertices in $S_{1}\cup
S_{2}-(A_{l-1}\cup A_{r+1}), i\neq j\neq k\in \{1,2,3\}$, then the
minimal cut-strategies of $\mathcal{P}(l,r)$ with
$\omega(G[\mathcal{P}(l,r)]/X)=1$ are sets of the form
$X'-(A_{l-1}\cup A_{r+1})=\{v: v\in (A_{1}\cup A_{2}\cup A_{3})-
(A_{1}\cap A_{2}\cap A_{3})-(A_{l-1}\cup A_{r+1}) \}$\\
[2mm]{\bf Case 3.} If $\mathcal{P}(l,r)$ has at least four
maximal cliques and we let $A_{1},A_{2},A_{3},A_{4},\\
\cdots, A_{k}$, $k\geq 4$, denote the maximal cliques in $\mathcal{P}(l,r)$.\\
Hence applying Lemma $3.6$ to $\mathcal{P}(l,r)$ implies that all
minimal cut-strategies of $\mathcal{P}(l,r)$ with
$\omega(G[\mathcal{P}(l,r)]/X)\geq 2$ are sets of the form
$X'-(A_{l-1}\cup A_{r+1})=\{v: 2\leq p\leq k-1, v\in
A_{p}-(S_{p-1}\cup S_{p})-(A_{l-1}\cup A_{r+1})\ , or\ v\in
S_{p}-X-(A_{l-1}\cup A_{r+1})\, where\ 2\leq p\leq k-1, X=S_{1}\cup
S_{k-1}\cup (A_{1}\cap A_{2}\cap A_{3})\cup(A_{k-2}\cap A_{k-1}\cap
A_{k}),\ or\ v\in A_{1}-(S_{1}\cup S_{2})-(A_{l-1}\cup A_{r+1})\
and\ it\ is\ adjacent\ to\ all\ vertices\\ in\ S_{1}\cup S_{2},\ or\
v\in A_{k}-(S_{k-2}\cup S_{k-1})-(A_{l-1}\cup A_{r+1})\ and\ it\ is\
adjacent\ to\ all\\ vertices\ in\ S_{k-2}\cup S_{k-1} \}$, if there
exists no $S_{i}$ and $S_{j}$, $1\leq i\neq j\leq k-1$, such that
$S_{i}\subseteq S_{j}$. Otherwise, if there exist $S_{i}$ and
$S_{j}$, $1\leq i\neq j\leq k-1$, such that $S_{i}\subseteq S_{j}$,
then $X'=\{v: 1\leq p\leq k, v\in A_{p}-S_{j}-(A_{l-1}\cup
A_{r+1}),\ and\ it\ is\ adjacent\ to\ all\ vertices\ in\ S_{j}$\}.\\
[2mm]For every $v\in V$ we define $l(v)=min\{k: v\in A_{k}\}$ and
$r(v)=max\{k: v\in A_{k}\}$. Then for all $l, r$ with $1\leq l\leq
r\leq t$ and for every component $C$ of $\mathcal{P}(l,r)$ holds
$C=\mathcal{P}(l(C),r(C))$ with $l(C)=min\{l(v): v\in C \}$ and
$r(C)=max\{r(v): v\in C \}$, i.e., $C$ is a piece.\\
[2mm]Now let $X=X'\cap \mathcal{P}(l,r)$ be a minimal cut-strategy
of $\mathcal{P}(l,r)$, $1\leq l\leq r\leq t$. Then it is easy to see
that graph $G[\mathcal{P}(l,r)/X]=G[\mathcal{P}(l,r)/X']$ is either
the disjoint union of $G[\mathcal{P}(l,p-1)]$ and
$G[\mathcal{P}(p+1,r)]$, or is the disjoint union of
$G[\mathcal{P}(l+1,l+1)]$ and $G[\mathcal{P}(l+2,r)]$ or is the
disjoint union of $G[\mathcal{P}(l,p)]$ and $G[\mathcal{P}(p+2,r)]$,
or is the disjoint union of $G[\mathcal{P}(l,p-1)]$,
$G[\mathcal{P}(l+1,p)]$ and $G[\mathcal{P}(p+1,r)]$, or is equal to
one of them (in case that $\mathcal{P}(l,p)=\emptyset$ or
$\mathcal{P}(p+1,r)=\emptyset$) or is $\emptyset$. Hence the set of
components of $G[\mathcal{P}(l,r)/X']$ is equal to the union of the
set of components of $G[\mathcal{P}(l,p)]$ and of the set of
components of $G[\mathcal{P}(p+1,r)]$ or is equal to one of these
sets. Therefore, all components of $G[\mathcal{P}(l,r)/X']$ are
pieces. \qed

From the definition of {\it piece} of $G$, we know that there have
essentially two different types of pieces in an interval graph. A
piece is called {\it complete} if it induces a complete graph and it
is called a {\it noncomplete} otherwise. It is obvious that pieces
$\mathcal{P}(l,l)$ are complete or $\emptyset$. Furthermore, a piece
$\mathcal{P}(l,r)$, $l<r$, may also be complete. And for every
complete piece induced graph $G[\mathcal{P}(l,r)]$, $l<r$, holds
$$S(G[\mathcal{P}(l,r)])=1\ \ \ \ \ \ \ (2)$$ \\
Furthermore, when piece $\mathcal{P}(l,r)$, $l<r$, has two or three
maximal cliques, we know that $S(G[\mathcal{P}(l,r)])=0$ or $1$ by
Theorems $3.5$ and $3.6$.

If there are at least four maximal cliques in it, the induced
subgraph $G[\mathcal{P}(l,r)]$, $1\leq l\leq r\leq t$, has minimal
cut-strategy $X$ with $\omega(G[\mathcal{P}(l,r)]/X)\geq 2$. So, for
every noncomplete piece $G[\mathcal{P}(l,r)]$, $1\leq l\leq r\leq
t$, having at least four maximal cliques, holds
$$S(G[\mathcal{P}(l,r)])=max\{\sum_{i=1}^{k}max\{S(G[P_{i}]),1\}-
|X'\cap \mathcal{P}(l,r)|\} \ \ \ \ \ \ \  (3)$$ where the maximum
is taken over all minimal cut-strategies $X'\cap \mathcal{P}(l,r)$,
with $\omega(G[\mathcal{P}(l,r)]/X)\geq 2$, of graph
$G[\mathcal{P}(l,r)]$ and $X'$ is a minimal cut-strategy of $G$,
$P_{1}, P_{2}, \cdots,P_{k}$ are the connected components of
$G[\mathcal{P}(l,r)/X]$.

Let $G$ be an interval graph. If $G$ is complete, then $S(G)=1$.
Otherwise the `dynamic programming on pieces' works as
follows:\\
[2mm]{\bf Step 1.} Compute a consecutive clique arrangement $A_{1},
A_{2}, \cdots, A_{t}$ of $G$, then compute $l(v)=min\{k: v\in
A_{k}\}$ and $r(v)=max\{k: v\in A_{k}\}$ for every $v\in V$,
and then compute all minimal cut-strategies.\\
[2mm]$(a)$ When $t=2$, $v\in X=\{v: v\in A_{1}-S_{1}\ and \ N(v)\cap
(A_{2}-S_{1})=\emptyset, or\ v\in A_{2}-S_{1}\ and\
N(v)\cap(A_{1}-S_{1})=\emptyset \}$.\\
[2mm]$(b)$ When $t=3$, $v\in X=\{v: v\in A_{i}-(S_{1}\cup
S_{2})-(A_{j}\cup A_{k}),\ i\neq j\neq k\in \{1,2,3\},\ and\ it\ is\
adjacent\ to\ all\ vertices\ in\ S_{1}\cup S_{2}\}$, or $v\in
X=\{v: v\in (A_{1}\cup A_{2}\cup A_{3})- (A_{1}\cap A_{2}\cap A_{3})\}$.\\
[2mm]$(c)$ When $t\geq4$, $v\in X=\{v: 2\leq p\leq t-1, v\in
A_{p}-(S_{p-1}\cup S_{p})\ , or\ v\in S_{p}-X\, where\ 2\leq p\leq
t-2, X=S_{1}\cup S_{t-1}\cup (A_{1}\cap A_{2}\cap
A_{3})\cup(A_{t-2}\cap A_{t-1}\cap A_{t}),or\ v\in A_{1}-(S_{1}\cup
S_{2})\ and\ it\ is\ adjacent\ to\ all\ vertices\ in\ S_{1}\cup
S_{2}, or\ v\in A_{t}-(S_{t-2}\cup S_{t-1})\ and\ it\ is\ adjacent\
to\ all\ vertices\ in\ S_{t-2}\cup S_{t-1} \}$, if there exists no
$S_{i}$ and $S_{j}$, $i\neq j$, such that $S_{i}\subseteq S_{j}$.
Otherwise, if there exist $S_{i}$ and $S_{j}$, $i\neq j$, such that
$S_{i}\subseteq S_{j}$, then $v\in X=\{v: 1\leq p\leq t, v\in
A_{p}-S_{j},\ and\ it\ is\ adjacent\ to\ all\ vertices\ in\ S_{j}$\}.\\
[2mm]{\bf Step 2.} For all $l,r$ with $1\leq l\leq r\leq t$
compute the vertex set $\mathcal{P}(l,r)$, mark $(l,r)$ `empty' if
$\mathcal{P}(l,r)=\emptyset$ and mark $(l,r)$ `complete' if
$\mathcal{P}(l,r)\neq\emptyset$ and $G[\mathcal{P}(l,r)]$ is a
complete induced graph.\\
[2mm]{\bf Step 3.} For all nonmarked tuples $(l,r)$ check whether
$G[\mathcal{P}(l,r)]$ is connected. If so, mark $(l,r)$
`noncomplete'. Else, mark $(l,r)$ `disconnected', and then compute
the components $P_{j}=\mathcal{P}(l_{j},r_{j})$, $1\leq j\leq k$, of
$G[\mathcal{P}(l,r)]$ and store
$(l_{1},r_{1}),(l_{2},r_{2}),\cdots,(l_{k},r_{k})$ in a linked
list with a pointer from $(l,r)$ to the head of this list.\\
[2mm]{\bf Step 4.} For all marked `noncomplete' tuples $(l,r)$,
$1\leq l\leq r\leq t$, compute the components
$P_{j}=\mathcal{P}(l_{j},r_{j})$, $1\leq j\leq k$, of
$G[\mathcal{P}(l,r)/v]$, where $v$ is a cut-strategy of
$G[\mathcal{P}(l,r)]$, and then check whether $\{v\}\cap
\mathcal{P}(l,r)$, is a minimal cut-strategy of
$G[\mathcal{P}(l,r)]$, and if so, mark $(v,l,r)$ `minimal', store
$(l_{1},r_{1}),(l_{2},r_{2}),\cdots,(l_{k},r_{k})$ in a linked list
with a pointer from $(v,l,r)$ to the head of this list
and it is obvious that $|\{v\}\cap \mathcal{P}(l,r)|=1$.\\
[2mm]{\bf Step 5.} For every pair $(l,r)$ marked `complete'
compute $S(G[\mathcal{P}(l,r)])$ according to $(2)$.\\
[2mm]{\bf Step 6.} For $d:=1$ to $t$ for $l:=1$ to $t-d$, if
$(l,l+d)$ is marked `noncomplete', compute
$S(G[\mathcal{P}(l,l+d)])$ according to Theorem $3.4$ if
$G[\mathcal{P}(l,l+d)]$ has two maximal cliques, according to
Theorem $3.6$ if $G[\mathcal{P}(l,l+d)]$ has three maximal cliques,
and according to $(3)$ when $G[\mathcal{P}(l,l+d)]$ has at least
four maximal cliques.\\
[2mm]{\bf Step 7.} Output $S(G)=S(G[\mathcal{P}(1,t)])$.

\noindent{\bf Theorem 3.9} {\it The above algorithm can compute
the neighbor-scattering number for interval graphs with time complexity
$O(n^{4})$.}\\
[2mm]{\bf Proof.} The correctness of this algorithm follows from
Theorem $2.2$ and lemma $3.6$. It is easy to see that steps
$1,2,5,7$ can be done in time $O(n^{4})$ in a straightforward
manner. In step $3$, testing connectedness and computing the
components can be done by an $O(n+m)$ algorithm for at most $n^{2}$
graphs $G[\mathcal{P}(l,r)]$. If $G[\mathcal{P}(l,r)]$ is
disconnected and $P_{j}$ is a component, then
$P_{j}=\mathcal{P}(l_{j},r_{j})$, $1\leq j\leq k$, with
$l_{j}=min\{l(v): v\in P_{j}\}$ and $r_{j}=max\{r(v): v\in P_{j}\}$
which can be computed in time $O(n)$. Hence, step $3$ can be done in
time $O(n^{4})$.

Step $4$ has to be executed for at most $n^{3}$ triples $(v,l,r)$
with $v\in V(G[\mathcal{P}(l,r)])$. If
$\mathcal{P}(l,r)/v\neq\emptyset$, then the components of
$G[\mathcal{P}(l,r)/v]$ are computed as indicated in the proof of
Lemma $3.7$ by using the marks of $(l,p-1)$ and $(p+1,r)$, or
$(l+1,l+1)$ and $(l+2,r)$, {\it etc.}, namely, if the mark is
`complete' or `noncomplete', then $(l,p-1)$ and $(p+1,r)$, or
$(l+1,l+1)$ and $(l+2,r)$, {\it etc.}, respectively, are stored and
if the mark is `disconnected', then the corresponding linked list is
added. Thus the linked list of $(v,l,r)$ can be computed in time
$O(n)$. As we know that $\{v\}\cap \mathcal{P}(l,r)$ is a minimal
cut-strategy of $G[\mathcal{P}(l,r)]$ if and only if $(a)$ or $(b)$
in Lemma $2.1$ holds. Because of the properties of a consecutive
clique arrangement it suffices to check that two components $P_{j}$
of $G[\mathcal{P}(l,p)]$ with the two largest values of $r_{j}$ and
the two components of $P_{j}$ of $G[\mathcal{P}(p+1,r)]$ with the
two smaller values of $l_{j}$ (if they exist). This can be done in
time $O(n)$. Hence step $4$ needs time $O(n^{4})$.

Step $6$ requires the evaluation of the right-hand side of $(3)$ for
at most $n^{2}$ pairs $(l,l+d)$. For every $v\in
V(G[\mathcal{P}(l,l+d)])$ and $(v,l,l+d)$ marked `minimal' the
components $P_{j}$ of $G[\mathcal{P}(l,l+d)/v]$ can be obtained in
time $O(n)$ from the linked list of $(v,l,l+d)$. Each of the at most
$n$ values $S(G[P_{i}])$ can be determined in constant time by table
look-up since the neighbor-scattering numbers of smaller pieces are
already known. Thus $\sum_{i=1}^{k}max\{S(G[G[P_{i}]),1\}-|\{v\}\cap
\mathcal{P}(l,l+d)|$ can be evaluated in time $O(n)$. Consequently,
step $6$ of the algorithm can be done in time $O(n^{4})$.\\

\end{document}